\documentclass{aipproc}
\layoutstyle{8x11single}
\usepackage{amsmath,amsthm,amssymb}
\usepackage{multirow}
\usepackage{dcolumn}

\begin{document}

\title{Heavy hybrids in pNRQCD}


\author{Jaume Tarr\'us Castell\`a}{
 address={Physik-Department, Technische Universit\"at M\"unchen, \\ James-Franck-Str. 1, 85748 Garching, Germany},
 altaddress={Excellence Cluster Universe, Technische Universit\"at M\"unchen, \\ Boltzmann-str. 2, 85748, Garching, Germany},
 email={jaume.tarrus@tum.de}
}

\begin{abstract}
We report on the calculation of the Quarkonium Hybrid masses for the $cg\bar{c}$, $bg\bar{c}$ and $bg\bar{b}$ systems~\cite{hbrpaper}. A QCD analog of the Born-Oppenheimer approximation including the $\Lambda$--doubling terms has been used. The gluonic static energies are obtained combining the short range description from potential non--relativistic QCD (pNRQCD) and lattice QCD calculations. The effect $\Lambda$--doubling terms is to mix the contributions of the static energies to the hybrid states leading to the break of the degeneracy between spin symmetry multiplets with opposite parity. The spin symmetry multiplets with mixed contributions from different static energies have a lower mass with respect to the parity opposite that remains without mixed contributions. We compare with direct lattice calculations and discuss several experimental candidates. 
\end{abstract}

\maketitle

\section{Introduction}

During the past years experimental observations have revealed the existence of a large number of unexpected states close or above open flavor thresholds in the heavy quarkonium spectrum. Many of these state do no fit the standard quarkonium potential models and are called Exotics and labeled as $X$s, $Y$s and $Z$s. There is an ongoing significant amount experimental effort to study Exotic quarkonium: new states, production mechanisms, decays and transitions, precision and high statistics data, at Belle2, Babar, BESIII and LHCb. The number of this states has grown to more than a dozen states, see~\cite{Bodwin:2013nua,Brambilla:2014jmp} for a recent review.

Exotic quakonium states are particularly interesting because are candidates to non--traditional hadronic states, for example hadrons containing four constituent quarks or an exited gluon constituent. Charged exotic quarkonium should be necessarily described by an hypothesis including four constituent quarks, consisting of a heavy quark--antiquark pair together with a pair of light quarks. Many theoretical descriptions pictures involving four constituent quarks have been proposed to describe this states: meson molecules, diquarkonium, compact tetraquarks and hadro--quarkonium. The molecular approach in particular has been quite successful in explaining the states close to heavy meson thresholds, such as the $X(3872)$. It remains however, a significant amount of not well understood states.

A different approach is to consider quarkonium hybrids. Quarkonium hybrids consist of a heavy quark and an antiquark in a color octet configuration together with a gluonic excitation. Quarkonium hybrids are characterized by the vast dynamical difference between the slow and massive heavy quarks with the fast and massless gluons. Since the gluons move much faster than the heavy quarks it is expected that the gluonic fields adapt nearly instantaneously to the changes in the heavy quarks separation. 

This physical situation is analogous to that of molecules, where the nuclei play the role of the heavy degrees of freedom and the electrons that if the light degrees of freedom. To solve the Schr\"odinger equation for these systems the Born--Oppenheimer approximation was developed. It provides a method to approximately solve the Schr\"{o}dinger equation for the nuclei and electrons by exploiting the fact that the masses of the nuclei are much larger than the electron mass, and therefore the time scales for their dynamics are very different. 

There have been attempts to export the Born-Oppenheimer approximation to describe hybrid quarkonium~\cite{Juge:1997nc,Braaten:2014qka}. This can be done by defining the gluonic static energies in non-relativistic QCD (NRQCD) and computing them on the lattice. In the short distance limit between the two heavy quarks the static energies can be described using pNRQCD \cite{Brambilla:1999xf,Pineda:1997bj}. One can then use these static energies in the Schr\"{o}dinger equation for the heavy quarks to obtain the heavy quark--antiquark energy eigenstates and the corresponding wave functions, thus determining the hybrid masses.

Since the gluonic static energies depend on the heavy quark distance $r$, the kinetical operator of the heavy quarks ($H_{k}$) acts non--trivially on the gluonic states. Being more specific, the gluonic contributions to the angular piece of $H_{k}$ can be as important as the contributions from the heavy quarks, therefore they should not be neglected. The action of the angular derivative on the gluonic wave functions can be split into a part which leaves them invariant and a part which which mixes them. The latter can be neglected only when the energy eigenstates static energies are well separated. Since the gluonic wave functions can not be obtained, additional approximations have to be made to estimate the contribution of the angular derivatives of $H_{k}$ acting on them.

\section{Gluonic static energies}

The static energies can be classified by the symmetries of the system. The symmetry transformations which leave a system of two static particles invariant form the group $D_{\infty h}$, which is the symmetry group of a cylinder. The basic transformations are rotations around the cylinder axis $R(\alpha)$, reflections across a plane including the cylinder axis $M$ and space inversion $P$. All other transformations can be expressed as a combination of these. In the case of a system with a static particle and a static antiparticle, space inversion needs to be combined with charge conjugation to form a symmetry  $CP$. Note that all these symmetry transformations are understood to act on the light degrees of freedom only.

Representations need to fulfill the identities $R(\alpha)R(-\alpha)=1$, $R(2\pi n)=1$, $M^2=1$ and $CP^2=1$. So if we replace $R(\alpha)\to R(\Lambda\alpha)$ with $\Lambda\in\mathbf{Z}$, $M\to\sigma M$ and $P\to-\eta P$ with $\sigma,\eta\in\{-1,+1\}$, we obtain another irreducible representation. The two-dimensional representations $R(\Lambda\alpha)$ and $R(-\Lambda\alpha)$ are related by a similarity transform which leaves $M$ and $CP$ invariant, so we can restrict $\Lambda$ for the two-dimensional case to have positive integer values. We will label the two components by $\lambda=\pm\Lambda$. Also $M$ and $-M$ are related by a similarity transform which leaves $R(\alpha)$ and $CP$ invariant, so a different sign under reflections does not lead to a different representation. The one dimensional representations are the same for any value of $\Lambda$, so we assign them $\Lambda=0$. In this case $M$ and $-M$ are not related and lead to different irreducible representations. General irreducible representations can then be labeled $\Lambda_\eta^\sigma$, where the superscript $\sigma$ only appears for $\Lambda=0$. Traditionally, the letters $g$ and $u$ are used for $\eta=+1$ and $\eta=-1$, and for $\Lambda=0,1,2,\dots$ uppercase Greek letters $\Sigma,\Pi,\Delta,\dots$ are used that correspond to the $s,p,d,\dots$ atomic orbital labels.

The static energies can be labeled by $\Lambda$, $\eta$ and $\sigma$. Usually, the symbol $\Lambda_\eta^\sigma$ is reserved for the lowest static energy of those quantum numbers, the next higher static energy of the same quantum numbers is written ${\Lambda_\eta^\sigma}'$ and so on. Alternatively, one can introduce another quantum number $n$ to distinguish between different static energies that correspond to the same irreducible representation by writing them as $n\Lambda_\eta^\sigma$.

The static energies for heavy quark--antiquark pairs, which are labeled by these quantum numbers, have been computed in quenched lattice QCD by Juge, Kuti and Morningstar~\cite{Juge:1997nc,Juge:2002br}. Additional lattice simulations were carried out by Bali and Pineda in~\cite{Bali:2003jq} focusing on the short range static energies for the $\Pi_u$ and $\Sigma^-_u$ potentials. 


\begin{figure}[ht]
 \centering
 \includegraphics[width=0.45\linewidth]{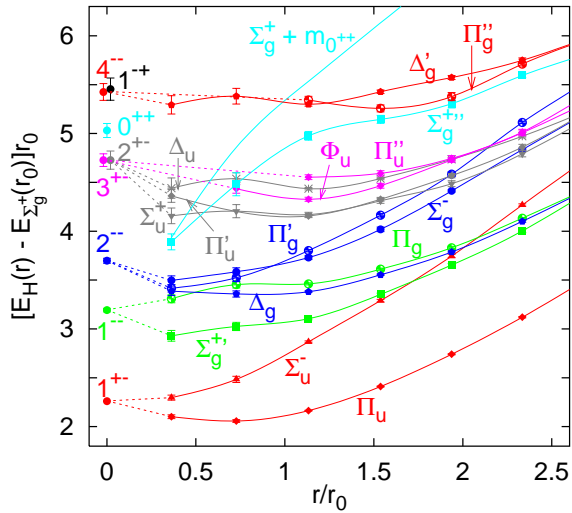}
 \caption{The lowest hybrid static energies~\cite{Juge:2002br} and gluelump masses~\cite{Foster:1998wu} in units of $r_0$. The absolute values have been fixed such that the ground state $\Sigma_g^+$ static energy (not displayed) is zero at $r_0$.}
 \label{hybstat}
\end{figure}

In Fig.~\ref{hybstat} the lattice data from Ref.~\cite{Juge:2002br} is plotted and compared with the gluelump spectrum, computed also on the lattice, of Ref.~\cite{Foster:1998wu}. We can see that the two lowest lying hybrid static energies are the $\Pi_u$ and $\Sigma^-_u$ states and they clearly tend to form a degenerate multiplet in the short range. The $\Pi_g-\Sigma_g^{+\prime}$, $\Delta_g-\Sigma^-_g-\Pi_g'$ and $\Delta_u-\Pi_u'-\Sigma^+_u$ multiplets are also expected to be degenerate in the short range~\cite{Brambilla:1999xf}.

\section{Orbital wave functions and $J^{PC}$ eigenstates} \label{owf}

The eigenstates of the static Hamiltonian, which we will write as $|n,\lambda,\eta,\sigma\rangle$, can be combined with orbital wave functions to form $J^{PC}$ eigenstates. First we construct eigenstates of the angular momentum operator $\mathbf{L}=\mathbf{K}+\mathbf{L}_{Q\overline{Q}}$, where $\mathbf{K}$ stands for the total angular momentum operator for the light degrees of freedom and $\mathbf{L}_{Q\overline{Q}}$ denotes the angular momentum operator of the heavy quark--antiquark pair. Note that the static states $|n,\lambda,\eta,\sigma\rangle$ implicitly depend on the distance $\mathbf{r}$ between the quarks. As an element of the kinetic term for the heavy quarks, $\mathbf{L}_{Q\overline{Q}}$ acts on this implicit dependence in $|n,\lambda,\eta,\sigma\rangle$.

The construction of the eigenstates of $\mathbf{L}$ can be done exactly without neglecting any of these effects. Let us define $\mathbf{r}=r\mathbf{\hat{n}}$. It can be shown that the operators $L^2$, $L_3$ and $\mathbf{\hat{n}}\cdot\mathbf{K}$, whose eigenvalue is $\lambda$, all commute with each other and with the static Hamiltonian. So we can define simultaneous eigenstates of those operators by writing
\begin{equation}
 |l,m_l;n,\lambda,\eta,\sigma\rangle=\int d\Omega\,|n,\lambda,\eta,\sigma\rangle\,v_{l,m_l}^\lambda(\theta,\phi)\,.
\end{equation}
The orbital wave functions $v_{l,m}^\lambda(\theta,\phi)$ can be found in textbooks such as~\cite{LandauLifshitz}.

The heavy quark spin $\mathbf{S}$ can be added to the angular momentum $\mathbf{L}$ to give the total angular momentum $\mathbf{J}=\mathbf{L}+\mathbf{S}$ using the usual Clebsch-Gordan coefficients. The states obtained in this way transform in the following way under space inversion and charge conjugation:
\begin{equation}
\begin{split}
 P|j,m,s,l;n,&\lambda,\eta,\sigma\rangle=\sigma_P(-1)^{l+1}|j,m,s,l;n,-\lambda,\eta,\sigma\rangle\,,\\
 C|j,m,s,l;n,&\lambda,\eta,\sigma\rangle=\eta\sigma_P(-1)^{l+s}|j,m,s,l;n,-\lambda,\eta,\sigma\rangle\,.
\end{split}
\end{equation}
These transformations turn $\lambda$ into $-\lambda$, because $\mathbf{\hat{n}}$ changes sign and $\mathbf{K}$ remains invariant. We can combine two $\lambda=\pm\Lambda$ states ($\Lambda\geq1$) to give eigenstates of $P$ and $C$ with eigenvalues $\epsilon$ and $\epsilon\,\eta\,(-1)^{s+1}$ respectively:
\begin{equation}
\begin{split}
|j,m,s,l;n,&\Lambda,\eta,\epsilon\rangle=\frac{1}{\sqrt{2}}\left(|j,m,s,l;n,\Lambda,\eta\rangle+\epsilon\sigma_P(-1)^{l+1}|j,m,s,l;n,-\Lambda,\eta\rangle\right)\,.
\end{split}
\end{equation}

\section{Radial wave functions and Schr\"{o}dinger equations} \label{rwfse}

A hybrid state consist of an angular wave function, given by $J^{PC}$ eigenstates of the previous section and a radial wave function. The last step in the Born-Oppenheimer approximation is to obtain the radial wave functions and energy eigenvalues from a Schr\"{o}dinger equation. If we act with the kinetic term on a hybrid state, it will affect both the radial wave function and the $J^{PC}$ eigenstate. 

The kinetic term can be split into radial and angular derivatives. The contribution of the radial derivative acting on the static states (which are not known explicitly) can be neglected, because it effectively acts as a correction to the static energies suppressed by the heavy quark mass. The angular term $\mathbf{L}_{Q\overline{Q}}^2$ acts on both the static states and the heavy quark angular wave functions. This can be worked out as follows. We will use projections onto the set of orthonormal vectors associated with spherical coordinates: $(\mathbf{\hat{\theta}},\mathbf{\hat{\phi}},\mathbf{\hat{n}})$, where $\mathbf{\hat{\theta}}=\partial_\theta\mathbf{\hat{n}}$ and $\mathbf{\hat{\phi}}=\sin^{-1}\theta\,\partial_\phi\mathbf{\hat{n}}$. With these we can write $\mathbf{K}=K_\theta\mathbf{\hat{\theta}}+K_\phi\mathbf{\hat{\phi}}+K_n\mathbf{\hat{n}}$ and $K_\pm=K_\theta\pm iK_\phi$. Then
\begin{equation}
 \mathbf{L}_{Q\overline{Q}}^2=(\mathbf{L}-\mathbf{K})^2=\mathbf{L}^2-2\mathbf{L}\cdot\mathbf{K}+\mathbf{K}^2=\mathbf{L}^2-2K_n^2+\mathbf{K}^2-L_-K_+-L_+K_-\,.
 \label{LQQbar}
\end{equation}
The first two are diagonal on our basis of states and give exactly $l(l+1)$ and $-2\Lambda^2$. The last three operators are not diagonal on our basis of states and can not be computed directly without the gluonic wave functions, which are unknown. The last two terms, $L_-K_+-L_+K_-$, raise or lower $\Lambda$ and $\mathbf{L}$ by $1$.

The kinetic term $\mathbf{L}_{Q\overline{Q}}^2$ is divided by $r^2$, therefore it should be most important in the region where $r$ is close to zero. In this limit there is no preferred direction for the static system, so the symmetry group $D_{\infty h}$ is extended to $O(3)\times C$, whose irreducible representations are $K^{PC}$ multiplets. The irreducible representations of $D_{\infty h}$ turn into components of these multiplets and the energy eigenvalues of the corresponding static states become degenerate in this limit. An approximation of $\mathbf{K}^2-L_-K_+-L_+K_-$ can be obtained on the $r\to0$ limit. We can take $\langle\mathbf{K}^2\rangle=k(k+1)$ with $k$ from the $K^{PC}$ multiplet which contains the respective $n\Lambda_\eta^\sigma$ representation. For the three static states with the lowest energies, $\Sigma_g^+$, $\Pi_u$ and $\Sigma_u^-$, these would be $0^{++}$, $1^{+-}$ and $1^{+-}$ respectively. 

The corrections from the last two terms in eq.~\eqref{LQQbar} lead to an effect called $\Lambda$-doubling~\cite{LandauLifshitz}. For $\Lambda\geq1$ there are always two opposite parity states which are mass degenerate if the mixing terms are ignored. This degeneracy is lifted by the corrections from those terms, which act differently on the different parity states. It can be shown that
\begin{equation}
\begin{split}
&\left\langle l;n',\lambda\pm1\left|L_\mp K_\pm\right|l;n,\lambda\right\rangle=\sqrt{l(l+1)-\lambda(\lambda\pm1)}\left\langle n',\lambda\pm1\left|K_\pm\right|n,\lambda\right\rangle\,.
\end{split}
\end{equation}
So far we did not specify how the quantum number $n$ should be assigned for static states with different $\Lambda_\eta^\sigma$ quantum numbers. It is convenient to define $n$ such that different static states with the same $r\to0$ limit, i.e.~which approach the same $K^{PC}$ multiplet and the same energy eigenvalue, also have the same $n$ quantum number. In the $r\to0$ limit $K_\pm$ can only facilitate $\lambda$ transitions within the same $K^{PC}$ multiplet. Thus we get
\begin{equation}
 \left\langle n',\lambda\pm1\left|K_\pm\right|n,\lambda\right\rangle\approx\sqrt{k(k+1)-\lambda(\lambda\pm1)}\,\delta_{n'n}\,.
\end{equation}
The mixing effects of $L_-K_++L_+K_-$ can be fully incorporated by working with coupled Schr\"{o}dinger equations. To this end, we define the hybrid states as superpositions of different $\Lambda$ states with the same $n$, i.e.~that approach the same $K^{PC}$ multiplet in the $r\to0$ limit:
\begin{equation}
\begin{split}
 |N;j,m,s,l;&n,\eta,\epsilon\rangle=\sum_\Lambda\int dr\,|j,m,s,l;n,\Lambda;\eta,\epsilon\rangle\,\frac{\psi_N^\Lambda(r)}{r}\,,
\end{split}
\end{equation}
where the sum over $\Lambda$ runs from $0$ or $1$, depending on whether a $\Lambda=0$ state with the same $\epsilon$ exists for that $n$, to the smaller value of either $k$ or $l$. The radial wave functions $\psi_N^\Lambda(r)$ are solutions to the coupled Schr\"{o}dinger equations~\cite{hbrpaper}
\begin{equation}
\begin{split}
& \left(-\frac{\partial_r^2}{2\mu}+\frac{C_{lk}^{\Lambda\,0}}{2\mu r^2}+E_{n\Lambda_\eta^\sigma}(r)\right)\psi_N^\Lambda(r)-\frac{C_{lk}^{\Lambda\,-}}{2\mu r^2}\psi_N^{\Lambda-1}(r)-\frac{C_{lk}^{\Lambda\,+}}{2\mu r^2}\psi_N^{\Lambda+1}(r)=E_N\psi_N^\Lambda(r)\,,
\end{split}
\end{equation}
where 
\begin{align}
 C_{lk}^{\Lambda\,0}&=l(l+1)-2\Lambda^2+k(k+1)\,,\\
 C_{lk}^{\Lambda\,\pm}&=\sqrt{l(l+1)-\Lambda(\Lambda\pm1)}\sqrt{k(k+1)-\Lambda(\Lambda\pm1)}\,,\\
 C_{lk}^{1\,-}&=C_{lk}^{0\,+}=\sqrt{2}\sqrt{l(l+1)}\sqrt{k(k+1)}\,.
\end{align}
Lattice calculations indicate that the lowest static energies above the ground state $\Sigma_g^+$ belong to states in the $\Sigma_u^-$ and $\Pi_u$ representations. Figure~\ref{latdatpot} shows that they become degenerate in the $r\to0$ limit. If there are no other static states with the same small $r$ behavior, like in this case, then these $\Sigma_u^-$ and $\Pi_u$ states need to form a $K^{PC}$ multiplet with quantum numbers $1^{+-}$ in the $r\to0$ limit. We will set $n=1$ for them. The spin symmetry multiplets one can construct with these $\Sigma_u^-$ and $\Pi_u$ states are shown in Tables~\ref{meth1} and \ref{meth2}.

\section{Hybrid static energies in pNRQCD}

In this section we study how pNRQCD can be used to describe the hybrid static energies computed on the lattice. Quarkonium systems are non--relativistic bound states in which a natural hierarchy of scales arises $m\gg \mathbf{p}_Q \gg E_b$, where $m$ is the heavy quark mass, $\mathbf{p}_Q$ is the typical relative three--momentum and $E_b$ the typical bound state energy scale. We can exploit this scale hierarchies by building an Effective Field Theory (EFT). pNRQCD is an EFT that describes quarkonium systems at $E_b$ scale. Depending of the relative position of the $\Lambda_{QCD}$ scale in relation to $\mathbf{p}_Q$ a weakly and a strongly coupled versions of pNRQCD can be formulated. In the short distance limit $1/r\sim  \mathbf{p}_Q \gg \Lambda_{QCD}$ the behavior of the static energies can be studied using weakly--coupled pNRQCD. At leading order in the multipole expansion the singlet decouples from the octet and gluons. The octet is still coupled to gluons of energy $\sim \Lambda_{QCD}$. The singlet then corresponds to the lowest static energy, with quantum numbers $\Sigma_g^+$. Gluelumps are defined as gluonic field configurations in the presence of an adjoint source
\begin{equation}
H\left(\mathbf{R},\mathbf{r},t\right)=H^a\left(\mathbf{R},t\right)\mathcal{O}^{a\dagger}\left(\mathbf{R},\mathbf{r},t\right)\,,
\end{equation}
where $\mathcal{O}^a$ is the octet field and $H^a$ is the gluonic operator. In the short distance limit the spectrum of gluonic excitations corresponds to the gluelumps. All the $H^a$ operators up to dimension 3, classified according to their rotational quantum number $K$ and the representation of $D_{\infty h}$ they correspond to, can be found in Table~\ref{tab3}.
\begin{table}[htb]
\centerline{
\begin{tabular}{c|c|c}
$\Lambda_\eta^\sigma$   & $K^{PC}$  & $H^a$ \\ \hline
$\Sigma_u^-$            & $1^{+-}$ & ${\bf r}\cdot{\bf B} \;,{\bf r}\cdot({\bf D}\times {\bf E})$   \\ 
$\Pi_u$                 & $1^{+-}$ & ${\bf r}\times{\bf B}\;,{\bf r}\times({\bf D}\times {\bf E})$  \\
$\Sigma_g^{+\, \prime}$ & $1^{--}$ & ${\bf r}\cdot{\bf E} \;, {\bf r}\cdot({\bf D}\times {\bf B})$  \\
$\Pi_g$                 & $1^{--}$ & ${\bf r}\times{\bf E}\;,{\bf r}\times({\bf D}\times {\bf B}) $ \\
$\Sigma_g^-$            & $2^{--}$ & $({\bf r}\cdot {\bf D})({\bf r}\cdot {\bf B}) $                \\
$\Pi_g^{\prime}$        & $2^{--}$ & ${\bf r}\times(({\bf r}\cdot{\bf D}) {\bf B}+{\bf D}({\bf r}\cdot{\bf B}))$  \\
$\Delta_g$              & $2^{--}$ & $({\bf r}\times {\bf D})^i({\bf r}\times{\bf B})^j+({\bf r}\times{\bf D})^j({\bf r}\times{\bf B})^i$ \\
$\Sigma_u^{+}$          & $2^{+-}$ & $({\bf r}\cdot {\bf D})({\bf r}\cdot {\bf E})$                  \\
$\Pi_u^{\prime}$        & $2^{+-}$ & ${\bf r}\times(({\bf r}\cdot{\bf D}) {\bf E}+{\bf D}({\bf r}\cdot{\bf E})) $ \\
$\Delta_u$              & $2^{+-}$ & $({\bf r}\times {\bf D})^i({\bf r}\times{\bf E})^j+({\bf r}\times{\bf D})^j({\bf r}\times{\bf E})^i$ \\ \hline
\end{tabular}}
\caption{Gluonic excitation operators in pNRQCD up to dimension 3. Different projections of the same fields correspond to different $D_{\infty h}$ representations.}
\label{tab3}
\end{table}
The hybrid static energy spectrum reads
\begin{equation}
E_H=2 m+V_H\,,
\label{hse}
\end{equation}
where the hybrid potential $V_H$ is given by the correlators of the gluelump operators for large distances in time
\begin{equation}
\begin{split}
&V_H= \\
&\lim_{T \to \infty}\frac{i}{T}\log\left\langle H^a(T/2)\mathcal{O}^a(T/2)H^b(-T/2)\mathcal{O}^b(-T/2)\right\rangle\,.
\end{split}
\end{equation}
Up to next--to--leading order in the multipole expansion
\begin{equation}
V_H(r)=V_o(r)+\Lambda_H+b_H r^2\,,
\label{hybridpot}
\end{equation}
where $V_o(r)$ is the octet potential, which can be computed in perturbation theory, and $\Lambda_H$ corresponds to the gluelump mass
\begin{equation}
\begin{split}
&\Lambda_H= \\
&\lim_{T \to \infty}\frac{i}{T}\log\left\langle H^a(T/2)\phi^{adj}_{ab}(T/2,-T/2)H^b(-T/2)\right\rangle\,,
\end{split}
\end{equation}
where $\phi^{adj}(T/2,-T/2)$ is a Wilson line.

The constant $\Lambda_H$ depends in general on the particular operator $H^a$, however, it is the same for operators corresponding to different projections of the same gluonic operators. From Table~\ref{tab3} we can see that in the short distance limit the $\Pi_u-\Sigma_u^{-}$, $\Pi_g-\Sigma_g^{+\prime}$, $\Delta_g-\Sigma^-_g-\Pi_g'$ and $\Delta_u-\Pi_u'-\Sigma^+_u$ multiplets should be degenerate. The gluelump mass is a non--perturbative quantity and has been determined on the lattice~\cite{Bali:2003jq,Foster:1998wu,Marsh:2013xsa}.

We work in the Renormalon Subtracted (RS) scheme \cite{Bali:2003jq,Pineda:2001zq}. In this scheme the convergence of the octet potential is improved. Note that when working in the RS scheme for the octet potential and gluelump mass, the quark mass in the hybrid static energy also has to be taken in the RS scheme. We have used the RS octet potential $V^{RS}_o(r)$ up to order $\alpha^3_s$ in perturbation theory and $\Lambda^{RS}_H$ at the subtraction scale $\nu_f=1$~GeV. The values of the heavy quark and the $1^{+-}$ gluelump mass in the RS scheme at $\nu_f=1$~GeV are: $m^{RS}_c=1.477(40)$~GeV, $m^{RS}_b=4.863(55)$~GeV and $\Lambda_H^{RS}=0.87(15)$~GeV~\cite{Bali:2003jq,Pineda:2001zq}.

The next--to--leading order corrections to the hybrid static energies are proportional to $r^2$ due to the multipole expansion and rotational invariance. The specific proportionality constant depends on non--perturbative dynamics and is not known, in the present work we are going to fix it through a fit to the lattice data for the static energies. We are going to consider that this term takes different values for hybrid static energies corresponding to different representations of $D_{\infty h}$, thus breaking the degeneracy of the short range pNRQCD description of the $\Pi_u$ and $\Sigma^-_u$ static energies, in the leading order of the multipole expansion.

\section{Hybrid masses from the Schr\"odinger equation} \label{mrsts}

The hybrid masses can be obtained as the energy eigenvalues of Schr\"odinger equations with the two lowest lying hybrid static energies ($\Pi_u$ and $\Sigma_u^-$) as potentials. In order to obtain the coefficients of the quadratic terms $b_H$ of the potentials~\eqref{hybridpot} we have used the available lattice data on the hybrid static energies. From Juge, Kuti and Morningstar we have used the lattice data of Ref.~\cite{Juge:2002br}, and from Bali and Pineda we have used the continuum extrapolated data presented in Ref.~\cite{Bali:2003jq}.

These two sources of lattice data have different energy offsets that must be taken into account. We have extracted the energy offsets from both sets of lattice data with respect to the theoretical hybrid potential by fitting the function
\begin{equation}
\mathcal{V}(r)=V_o^{RS}+c+b_H r^2\,,
\label{potfit}
\end{equation}
with $c$ and $b_H$ as free parameters. We have performed a joint fit of both potentials of the form~\eqref{potfit} to the lattice data of both groups, restricting the value of $c$ to be the same for both potentials but different for each group and, conversely, restricting the value of $b_H$ to be the same for both groups but different for each potential.

\begin{figure}[ht]
 \centering
 \includegraphics[width=0.4\linewidth]{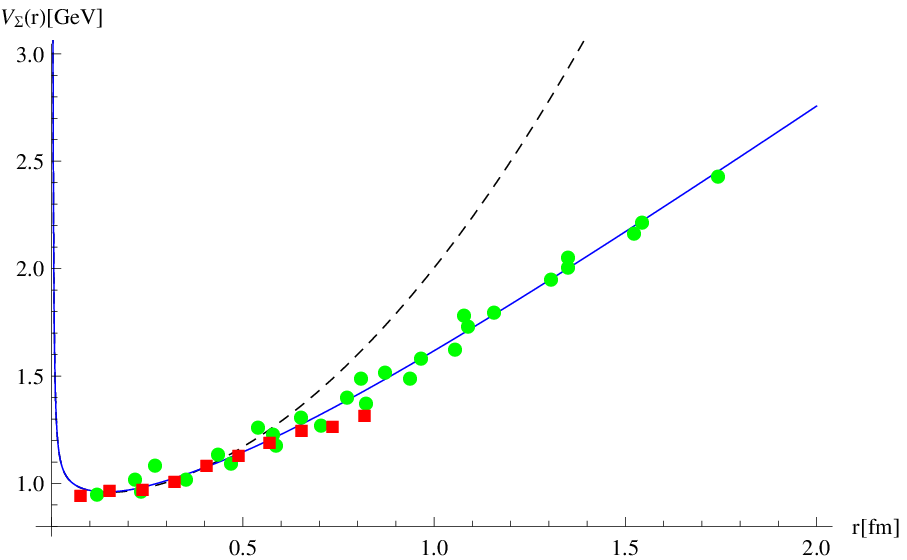}
 \includegraphics[width=0.4\linewidth]{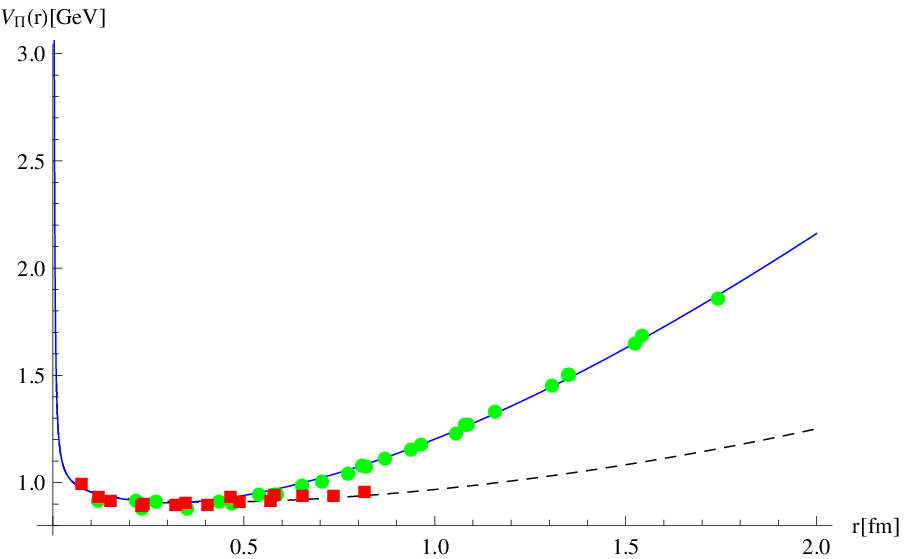}
 \caption{Lattice data from Bali and Pineda~\cite{Bali:2003jq} is represented by red squares, the data from Juge, Kuti and Morningstar~\cite{Juge:2002br} is represented by the green dots. In the left (right) figure we have plotted the data corresponding to the $\Sigma^-_u$ ($\Pi_u$). The lattice data has been corrected by the offsets of~\eqref{offsets}. The black dashed line corresponds to $V^{(0.5)}$, the blue continuous line to $V^{(0.25)}$.}
 \label{latdatpot}
\end{figure}

The pNRQCD description of the hybrid static energy of (\ref{hybridpot}) is only valid up to $r \gtrsim 1/\Lambda_{QCD}$. Taking perturbation theory up to its limit of validity, we have fitted (\ref{potfit}) to lattice data in the range of $r=0-0.5$~fm. We obtain the following offsets for the two lattice data sources 
\begin{equation}
c_{\text{BP}}=0.105\,{\rm GeV} ,\, c_{\text{KJM}}=-0.471\,{\rm GeV}\,,
\label{offsets}
\end{equation}
and the values for the coefficient of the quadratic term are
\begin{equation}
b^{(0.5)}_{\Sigma}=1.112\,{\rm GeV/fm^2},\, b^{(0.5)}_{\Pi}=0.110\,{\rm GeV/fm^2}\,.
\label{bfit05}
\end{equation}
The potentials obtained from using the coefficients of the quadratic terms of~\eqref{bfit05} in eq.~\eqref{hybridpot} will be called $V_{\Pi_u}^{(0.5)}$ and $V_{\Sigma^-_u}^{(0.5)}$ (corresponding to the $\Pi_u$ and $\Sigma^-_u$ configurations). We have plotted $V_{\Pi_u}^{(0.5)}$ and $V_{\Sigma^-_u}^{(0.5)}$ in Fig.~\ref{latdatpot} with the lattice data corrected for the different offsets using the values from~\eqref{offsets}. 
We have solved the coupled Schr\"odinger equations with the potentials $V^{(0.5)}$ and the RS heavy quark masses~\cite{hbrpaper}. The results are displayed in Table~\ref{meth1}. The largest source of uncertainties for the hybrid masses lies in the RS gluelump mass, which is known with an uncertainty of $\pm0.15$~GeV.
\begin{table}
\footnotesize{
 \begin{tabular}{||c|c||c|c|c|c||c|c|c|c||c|c|c|c||}
 \hline
 \multirow{2}{*}{\text{Multiplet}} & \multirow{2}{*}{$J^{PC}$} & \multicolumn{4}{c||}{$c\bar{c}$} & \multicolumn{4}{c||}{$b\bar{c}$} & \multicolumn{4}{c||}{$b\bar{b}$}\\
 \cline{3-6}\cline{7-10}\cline{11-14}
 & & $m_H$ & $\langle1/r\rangle$ & $\langle E_{k}\rangle$ & $P_\Pi$ & $m_H$ & $\langle1/r\rangle$ & $\langle E_{k}\rangle$ & $P_\Pi$ & $m_H$ & $\langle1/r\rangle$ & $\langle E_{k}\rangle$ & $P_\Pi$ \\
 \hline\hline
 $H_1$ & \multirow{2}{*}{$\{1^{--},(0,1,2)^{-+}\}$} & 4.05 & 0.29 & 0.11 & 0.94 & 7.40 & 0.31 & 0.08 & 0.94 & 10.73 & 0.36 & 0.06 & 0.95 \\
 \cline{3-14}
 $H_1'$ & & 4.23 & 0.27 & 0.20 & 0.91 & 7.54 & 0.30 & 0.16 & 0.91 & 10.83 & 0.36 & 0.11 & 0.92 \\
 \hline
 $H_2$ & \multirow{2}{*}{$\{1^{++},(0,1,2)^{+-}\}$} & 4.09 & 0.21 & 0.13 & 1.00 & 7.43 & 0.23 & 0.10 & 1.00 & 10.75 & 0.27 & 0.07 & 1.00 \\
 \cline{3-14}
 $H_2'$ & & 4.30 & 0.19 & 0.24 & 1.00 & 7.60 & 0.21 & 0.19 & 1.00 & 10.87 & 0.25 & 0.13 & 1.00\\
 \hline
 $H_3$ & $\{0^{++},1^{+-}\}$ &  4.69 & 0.37 & 0.42 & 0.00 & 7.92 & 0.42 & 0.34 & 0.00 & 11.09 & 0.50 & 0.23 & 0.00 \\
 \hline
 $H_4$ & $\{2^{++},(1,2,3)^{+-}\}$ & 4.17 & 0.19 & 0.17 & 0.97 & 7.49 & 0.25 & 0.14 & 0.97 & 10.79 & 0.29 & 0.09 & 0.98 \\
 \hline
 $H_5$ & $\{2^{--},(1,2,3)^{-+}\}$ & 4.20 & 0.17 & 0.18 & 1.00 & 7.51 & 0.19 & 0.15 & 1.00 & 10.80 & 0.22 & 0.10 & 1.00 \\
 \hline
 \end{tabular}}
 \caption{Hybrid energies obtained from solving the Schr\"odinger equation for the $V^{(0.5)}$ potentials with the RS heavy quark masses. All values are given in units of GeV. The primes on the multiplets indicate the first excited state of that multiplet. $\langle E_{k}\rangle$ is the mean value of the kinetic energy. $P_{\Pi}$ is a measure of the probability to find the hybrid in a $\Pi_u$ configuration, thus it gives a measure of the mixing effects.}
 \label{meth1}
\end{table}
In Fig.~\ref{latdatpot} we can see that the potentials $V^{(0.5)}$ describe the lattice data well up to $r\lesssim 0.55-0.65$~fm which corresponds to $1/r \gtrsim 0.36-0.30$~GeV. If we compare these values with the results obtained from the $\langle1/r\rangle$ from Table~\ref{meth1} we can see that for the $c\bar{c}$ system just the ground state of the lowest energy hybrid multiplet can be considered to fall inside of the range of validity of the $V^{(0.5)}$ potentials. In the case of $b\bar{c}$ and $b\bar{b}$, both the ground and the first exited states of the lowest energy multiplet fall inside the validity range of the potential, in particular the $b\bar{b}$ states are well inside it. Also the $H_3$ multiplet has large enough $\langle1/r\rangle$, but fails to fulfill the following criterion for $c\bar{c}$ and $b\bar{c}$.

The multipole expansion requires that $\langle1/r\rangle \gg \langle E_{k}\rangle$. As long as this hierarchy is preserved we can expect the Schr\"odinger equation picture of the system to be consistent. The results in Table~\ref{meth1} show that the multipole expansion holds for the lower energy multiplets, in particular for the $b\bar{b}$ states. However, the results obtained from solving the Schr\"odinger equation for $V^{(0.5)}$ potentials show that many of the hybrid states lie beyond the range of validity expected for this potential.

We have constructed a second potential that includes as much information as possible from the long range lattice data. This potential, which we will call $V^{(0.25)}$, is defined as follows. For $r\leq 0.25$~fm we use the potential from (\ref{hybridpot}) with different $b_H$ factors for each of the low lying hybrid static energies $\Pi_u$ and $\Sigma^-_u$. The $b_H$ factors are obtained through a fit of the function (\ref{potfit}) for each potential to lattice data up to $r=0.25$~fm from both sources with the offsets of (\ref{offsets}). The quadratic term factors resulting from this fit are
\begin{equation}
b^{(0.25)}_{\Sigma}=1.246\,{\rm GeV/fm^2},\, b^{(0.25)}_{\Pi}=0.000\,{\rm GeV/fm^2}\,.
\label{bfit025}
\end{equation}
For $r\geq 0.25$~fm we have used a fit of the function
\begin{equation}
\mathcal{V}'(r)=\frac{a_1}{r}+\sqrt{a_2 r^2+a_3}+a_4\,,
\label{lattfit}
\end{equation}
to the lattice data corrected by the offsets of (\ref{offsets})~\cite{hbrpaper}. To ensure a smooth transition we have imposed continuity up to first derivatives. 

In Fig.~\ref{latdatpot} we have plotted the potentials $V^{(0.25)}$ together with the lattice data. The $V^{(0.25)}$ potentials do a good job reproducing the whole range of lattice data. We have solved the coupled Schr\"odinger equations with the potentials $V^{(0.25)}$ and the RS heavy quark masses~\cite{hbrpaper}. The results are displayed in Table~\ref{meth2}.
\begin{table}
\footnotesize{
 \begin{tabular}{||c|c||c|c|c|c||c|c|c|c||c|c|c|c||}
\hline
 \multirow{2}{*}{\text{Multiplet}} & \multirow{2}{*}{$J^{PC}$} & \multicolumn{4}{c||}{$c\bar{c}$} & \multicolumn{4}{c||}{$b\bar{c}$} & \multicolumn{4}{c||}{$b\bar{b}$}\\
 \cline{3-6}\cline{7-10}\cline{11-14}
 & & $m_H$ & $\langle1/r\rangle$ & $\langle E_{k}\rangle$ & $P_\Pi$ & $m_H$ & $\langle1/r\rangle$ & $\langle E_{k}\rangle$ & $P_\Pi$ & $m_H$ & $\langle1/r\rangle$ & $\langle E_{k}\rangle$ & $P_\Pi$ \\ \hline\hline
 $H_1$ & \multirow{2}{*}{$\{1^{--},(0,1,2)^{-+}\}$} & 4.15 & 0.42 & 0.16 & 0.82 & 7.48 & 0.46 & 0.13 & 0.83 & 10.79 & 0.53 & 0.09 & 0.86 \\ \cline{3-14}
 $H_1'$ & & 4.51 & 0.34 & 0.34 & 0.87 & 7.76 & 0.38 & 0.27 & 0.87 & 10.98 & 0.47 & 0.19 & 0.87 \\  \hline
 $H_2$ & \multirow{2}{*}{$\{1^{++},(0,1,2)^{+-}\}$} & 4.28 & 0.28 & 0.24 & 1.00 & 7.58 & 0.31 & 0.19 & 1.00 & 10.84 & 0.37 & 0.13 & 1.00 \\  \cline{3-14}
 $H_2'$ & & 4.67 & 0.25 & 0.42 & 1.00 & 7.89 & 0.28 & 0.34 & 1.00 & 11.06 & 0.34 & 0.23 & 1.00 \\  \hline
 $H_3$ & $\{0^{++},1^{+-}\}$ & 4.59 & 0.32 & 0.32 & 0.00 & 7.85 & 0.37 & 0.27 & 0.00 & 11.06 & 0.46 & 0.19 & 0.00\\
 \hline
 $H_4$ & $\{2^{++},(1,2,3)^{+-}\}$  & 4.37 & 0.28 & 0.27 & 0.83 & 7.65 & 0.31 & 0.22 & 0.84 & 10.90 & 0.37 & 0.15 & 0.87 \\
 \hline
 $H_5$ & $\{2^{--},(1,2,3)^{-+}\}$ & 4.48 & 0.23 & 0.33 & 1.00 & 7.73 & 0.25 & 0.27 & 1.00 & 10.95 & 0.30 & 0.18 & 1.00 \\
 \hline
 $H_6$ & $\{3^{--},(2,3,4)^{-+}\}$ & 4.57 & 0.22 & 0.37 & 0.85 & 7.82 & 0.25 & 0.30 & 0.87 & 11.01 & 0.30 & 0.20 & 0.89 \\
 \hline
 $H_7$ & $\{3^{++},(2,3,4)^{+-}\}$ & 4.67 & 0.19 & 0.43 & 1.00 & 7.89 & 0.22 & 0.35 & 1.00 & 11.05 & 0.26 & 0.24 & 1.00 \\
 \hline
 \end{tabular}}
 \caption{Hybrid energies obtained from solving the Schr\"odinger equation for the $V^{(0.25)}$ potentials. All values are given in units of GeV. The primes on the multiplets indicate the first excited state of that multiplet. $\langle E_{k}\rangle$ is the mean value of the kinetic energy. $P_{\Pi}$ is a measure of the probability to find the hybrid in a $\Pi_u$ configuration, thus it gives a measure of the mixing effects.}
 \label{meth2}
\end{table}
The states obtained with $V^{(0.25)}$ lie above the one ones obtained using $V^{(0.5)}$. The masses of the states with smaller sizes have a better agreement, since both potentials agree in the short range. Checking for the validity of the multipole expansion in the results of Table~\ref{meth2} we find that, for charmonium, just the ground state of the $H_1$ multiplet fulfills the hierarchy condition $\langle1/r\rangle \gg E_{kin}$ reasonably well. In the $b\bar{c}$ case the potential picture holds for the ground states of the $H_1$, $H_2$ and $H_3$ multiplets and works reasonably well for the first excited state of the $H_1$ multiplet. In bottomonium the multipole expansion holds for nearly all of the multiplets studied.

\begin{figure}
 \centering
\includegraphics[width=0.5\linewidth]{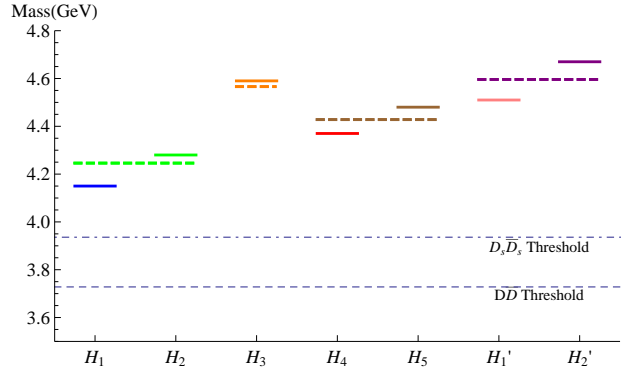}
\caption{Comparison of the hybrid multiplet masses in the charmonium sector obtained by Braaten \emph{et al.} \cite{Braaten:2014qka} with the results obtained using the $V^{(0.25)}$ potential. The Braaten \emph{et al.} results correspond to the dashed lines, while the solid lines corresponds to the results obtained using $V^{(0.25)}$. The degeneracy of the mass of the $H_{1/2}$ and $H_{4/5}$ multiplets in Braaten \emph{et al.} is broken by the introduction of the $\Lambda$-doubling terms.}
\label{bcc}
\end{figure}

In a recent paper, Braaten \emph{et al.}~\cite{Braaten:2014qka} followed a similar procedure to obtain the hybrid masses from the Born--Oppenheimer potentials computed on the lattice. They did not consider the operators that mix the contribution of the static energies in the Schr\"odinger equation. We have plotted the charmonium hybrid masses from \cite{Braaten:2014qka} together with our results obtained using the $V^{(0.25)}$ potential in Fig.~\ref{bcc}. The effect of the mixing terms is to break the degeneracy between the $H_1$ and $H_2$ multiplets as well as the $H_4$ and $H_5$ multiplets. The predicted $H_{1/2}$ mass from Braaten \emph{et al.} should be compared with our $H_2$ mass, since this multiplet is a pure $\Pi_u$ potential state. Similarly, their $H_{4/5}$ mass should be compared with our $H_5$ mass. The $H_3$ multiplet is a pure $\Sigma^-_u$ potential state in both approaches and can also be compared. We can see that there is a good agreement with our results from Table~\ref{meth2}, if we shift the masses by the difference in the $H_{2}$ state $\sim30$~MeV, then the other states agree within $40$~MeV. The mass shift of $30$~MeV should be accounted for through the different determination of the energies scale for the potentials. We can take the remaining $40$~MeV discrepancy between our results and those of~\cite{Braaten:2014qka} to be the uncertainty coming from the fitting of the potentials and the solution of the Schr\"odinger equation. Overall, comparing with the results from~\cite{Braaten:2014qka}, we can see that the effect of introducing the $\Lambda$--doubling breaking terms lowers the masses of the multiplets that have mixed contributions from two hybrid static energies.

\subsection{Identification of hybrids with experimental states}

\begin{figure}[ht]
 \centering
\includegraphics[width=0.8\linewidth]{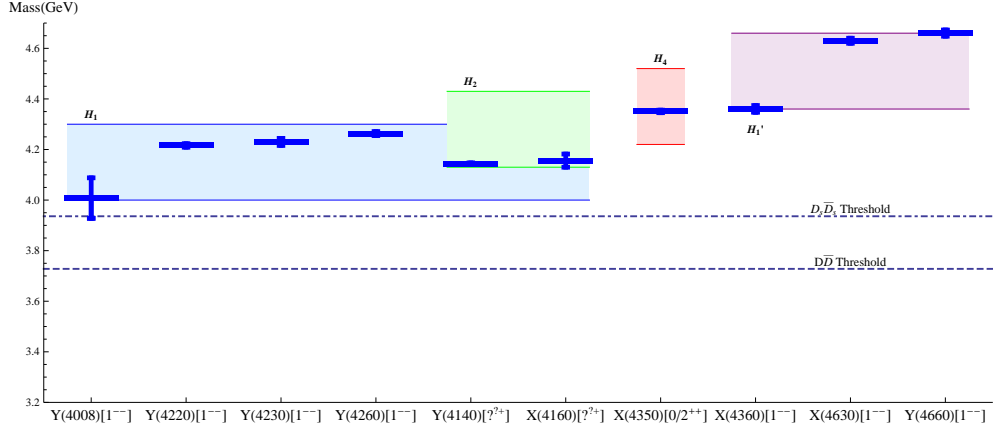}
\caption{Comparison of the experimental candidates masses for the charmonium sector with our results using the $V^{(0.25)}$ potential. The experimental states are plotted in solid blue lines with error bars corresponding to the average of the lower and upper mass uncertainties. Our results have been plotted in error bands corresponding to the gluelump mass uncertainty of $\pm0.15$~GeV.}
\label{exp}
\end{figure}

The list of candidates for heavy quark hybrids consists of the neutral heavy quark mesons above open flavor threshold. It is important to keep in mind that the main source of uncertainty of our results in section \ref{mrsts} is in the uncertainty of the gluelump mass $\pm0.15$~GeV. We have plotted the candidate experimental states in Fig.~\ref{exp}, except for the single one corresponding to the bottomonium sector, overlaid onto our results using the $V^{(0.25)}$ potential with error bands corresponding to the uncertainty of the gluelump mass.
Three $1^{--}$ states fall close to our mass for the charmonium hybrid from the $H_1$ multiplet, the $Y(4008)$, $Y(4230)$ and $Y(4260)$. The $1^{--}$ hybrid from the $H_1$ multiplet is a spin singlet state, and as such the decays to spin triplet products are suppressed by one power of the heavy quark mass due heavy quark spin symmetry. All these three candidate states decay to spin triplet charmonium, which in principle disfavors the hybrid interpretation. Nevertheless, there might be enough heavy quark spin symmetry violation to explain those decays. On the other hand the interpretation of this states as charmonium hybrids would make the decay into two $S$--wave open charm mesons forbidden \cite{Kou:2005gt}, which would explain why such decays have not been observed for the $Y(4260)$. Nevertheless the recent observation of the transition $Y(4260)\rightarrow X(3872)\gamma$~\cite{Ablikim:2013dyn} makes the identification of $Y(4260)$ as a hybrid highly unlikely.

The $Y(4220)$ is a narrow structure proposed in~\cite{Yuan:2013ffw} to fit the line shape of the annihilation processes $e^+e^-\rightarrow h_c\pi^+\pi^-$ observed by BESIII and CLEO-c experiments. It has mass is quite close to the one of the $H_1$ multiplet. As the previous states, it is a $1^{--}$ state that would be identified as a spin singlet hybrid. However, unlike the previous states, the $Y(4220)$ has been observed decaying to spin singlet quarkonium, which makes it a very good candidate to a charmonium hybrid. However the $Y(4220)$ falls very close to the $Y(4230)$ \cite{Ablikim:2014qwy} and it is possible that they are the same structure observed in different decay channels.

The $J^{PC}$ quantum numbers of the $Y(4140)$ and $Y(4160)$ have not yet been determined, however their Charge conjugation and mass suggest they can be candidates for the spin triplet $1^{-+}$ member of the $H_1$ multiplet, nevertheless their mass is also compatible within uncertainties with the spin singlet $1^{++}$ member of the $H_2$ multiplet. In the case of the $Y(4160)$, it decays into $D^* \bar D^*$  which favors a molecular interpretation of this state. If the $X(4350)$ turns out to be a $2^{++}$ state it can be a candidate for the spin singlet charmonium state of the $H_4$ multiplet, although its decay violates heavy quark spin symmetry. The three higher mass $1^{--}$ charmonia, the $X(4360)$, $X(4630)$ and $Y(4660)$ \footnote{It has been suggested that $X(4630)$ and are $Y(4660)$ might be actually the same particle.} have a mass that is compatible with the excited spin singlet member of the $H_1$ multiplet within uncertainties, although none of them falls very close to the central value. The $X(4360)$ and $Y(4660)$ decay into a spin triplet product which violates heavy quark spin symmetry.

There is so far only one bottomonium candidate for a hybrid state, the $Y_b(10890)$, which can be identified with the spin singlet $1^{--}$ state of the $H_1$ bottomonium hybrid multiplet. However, its decay to the $\Upsilon$ violates heavy quark spin symmetry, which is expected to be a good symmetry for bottomonium states.

\subsection{Comparison with direct lattice computations}

\begin{figure}
\centering
\includegraphics[width=0.8\linewidth]{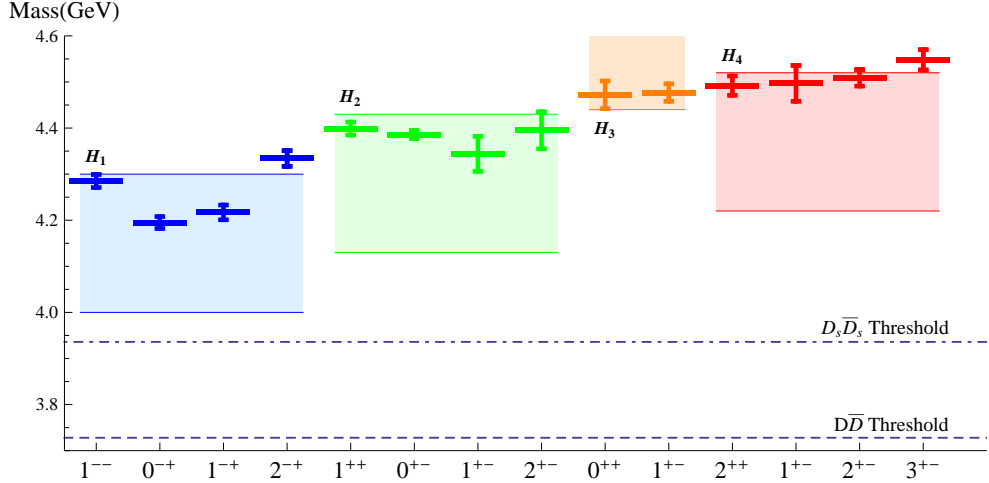}
\caption{Comparison of the results from direct lattice computations of the masses for charmonium hybrids Ref.~\cite{Liu:2012ze} with our results using the $V^{(0.25)}$ potential. The direct lattice mass predictions are plotted in solid lines with error bars corresponding the mass uncertainties. Our results have been plotted in error bands corresponding to the gluelump mass uncertainty of $\pm0.15$~GeV.}
\label{charmDLC}
\end{figure}

The spectrum of hybrids in the charmonium sector has been calculated by the Hadron Spectrum Collaboration~\cite{Liu:2012ze} using unquenched lattice QCD. The light quarks were given unphysically heavy masses equivalent to a pion mass of $\approx 400$~MeV. In Fig.~\ref{charmDLC} the results from Ref.~\cite{Liu:2012ze} have been plotted together with our results using the $V^{(0.25)}$ potential. The results from Ref.~\cite{Liu:2012ze} are given with the $\eta_c$ mass subtracted and are not extrapolated to the continuum limit. To compare with our results in Fig.~\ref{charmDLC} we use their results with the experimental value of $m_{\eta_c}=2.9837(7)$~GeV added. 
Comparing the spin averages of the masses of the hybrid states from Ref.~\cite{Liu:2012ze} to our results using the $V^{(0.25)}$ potential, we see that the masses obtained our masses are $0.1-0.14$~GeV lower except for the $H_3$ multiplet, which is $0.11$~GeV heavier. The mass splittings between the different multiplets are given in Table~\ref{mscomp}. We find a good agreement with the lattice data with our calculation using the $V^{(0.25)}$ potentials. In particular, the mass difference between $H_1$ and $H_2$, which in our calculation is directly related to the $\Lambda$-doubling effect, is in very close to our mass difference. The worst agreement is again found for the $H_3$ multiplet. It is interesting to note that the $H_3$ multiplet is the only one dominated by $\Sigma^-_u$ potential. The mass splittings between the different multiplets are given in Table~\ref{mscomp}. We find a good agreement with the lattice data with our calculation using the $V^{(0.25)}$ potentials. In particular, the mass difference between $H_1$ and $H_2$, which in our calculation is directly related to the $\Lambda$-doubling effect, is in very close to our mass difference. The worst agreement is again found for the $H_3$ multiplet.

\begin{table}
\centerline{
\begin{tabular}{c|c|c|c}
\text{Splitting}& Ref.~\cite{Liu:2012ze}   & $V^{(0.5)}$  & $V^{(0.25)}$ \\ \hline
$\delta m_{H_2-H_1}$  & $0.10$ & $0.04$ & $0.13$ \\
$\delta m_{H_4-H_1}$  & $0.24$ & $0.12$ & $0.22$ \\
$\delta m_{H_4-H_2}$  & $0.13$ & $0.08$ & $0.09$ \\
$\delta m_{H_3-H_1}$  & $0.20$ & $0.64$ & $0.44$ \\ 
$\delta m_{H_3-H_2}$  & $0.09$ & $0.60$ & $0.31$ \\ \hline
\end{tabular}}
\caption{Mass splittings between charmonium hybrid multiplets for the potentials $V^{(0.5)}$ and $V^{(0.25)}$ compared with the spin averages from the direct lattice calculation of Ref.~\cite{Liu:2012ze} in units of GeV.}
\label{mscomp}
\end{table}

\section{Conclusions} \label{conc}

We have reported~\cite{hbrpaper} on the computation of the heavy hybrid masses using a QCD analog of the Born-Oppenheimer approximation including the $\Lambda$-doubling terms by using coupled Schr\"odinger equations. The static energies have been obtained combining pNRQCD for short distances and lattice data for long distances. A large set of masses for spin symmetry multiplets for $c\bar{c}$, $b\bar{c}$ and $b\bar{b}$ hybrids has been obtained. The $\Lambda$-doubling effect breaks the degeneracy between opposite parity spin symmetry multiplets and has been found to lower the mass of the multiplets that get mixed contribution different static energies. The same pattern of $\Lambda$-doubling is observed in direct lattice calculations. Mass gaps between multiplets are in good agreement with the spin averaged direct lattice computation values, but the absolute values are shifted. Several experimental candidates for charmonium hybrids and one for bottomonium hybrids have been found, being $Y(4220)$ the most promising.

\begin{theacknowledgments}
We thank Colin Morningstar and Gunnar Bali for giving us acces to their static energies lattice data. This work has been supported by the DFG and the NSFC through funds provided to the Sino-German CRC 110 ``Symmetries and the Emergence of Structure in QCD''.
\end{theacknowledgments}

\bibliographystyle{aipproc}

\end{document}